\documentclass[journal, 9pt]{IEEEtran}

\usepackage[fleqn]{amsmath}
\usepackage{amssymb,graphicx,verbatim,enumitem,mdframed,mathrsfs}
\usepackage[numbers]{natbib}

\newtheorem{theorem}{\bfseries Theorem}

\newtheorem{lemma}{\bfseries Lemma}
\newtheorem{corollary}{\bfseries Corollary}
\newtheorem{definition}{\bfseries Definition}

\providecommand{\mnorm}[1]{|#1|}

\title{Information-Theoretic Privacy in \\Distributed Average Consensus} 

\author{Nirupam Gupta, Jonathan Katz, and Nikhil Chopra
\thanks{This work was supported by 
NSF Award \#ECCS1711554 and by
the Naval Air Warfare Center Aircraft Division, Pax River, MD, under contract N00421132M022.}
\thanks{Nirupam Gupta is with the Department of Computer Science, Georgetown University, Washington, DC 20057, USA (\small{\tt nirupam.gupta@georgetown.edu})}
\thanks{Jonathan Katz is with the Department of Computer Science, George Mason University, Fairfax, VA 22030, USA (\small{\tt katz@gmu.edu})}
\thanks{Nikhil Chopra is with the Department of Mechanical Engineering, University of Maryland, College Park, MD 20742, USA (\small{\tt nchopra@umd.edu})}
}

\begin{document}

\maketitle


\begin{abstract}                
We present a distributed average consensus protocol that preserves the privacy of agents' inputs. Unlike the differential privacy mechanisms, the presented protocol does not affect the accuracy of the output. It is shown that the protocol preserves the {\em information-theoretic} privacy of the agents' inputs against colluding passive adversarial (or honest-but-curious) agents in the network, if the adversarial agents do not constitute a vertex cut in the underlying communication network. This implies that we can guarantee information-theoretic privacy of all the honest agents' inputs against $t$ arbitrary colluding passive adversarial agents if the network is $(t+1)$-connected. The protocol is constructed by composing a distributed privacy mechanism that we propose with any (non-private) distributed average consensus algorithm.  
\end{abstract}



\section{\bfseries Introduction}
\label{sec:intro}
Distributed average consensus algorithms allow agents in a peer-to-peer network to reach consensus on the average value of their inputs \cite{jadbabaie2003coordination, olfati2007consensus, xiao2004fast}. Some common applications of distributed average consensus are; sensor fusion in a multi-sensor network~\cite{olfati2005distributed,olfati2005consensus}, decentralized support vector machine~\cite{forero2010consensus}, economic-dispatch problem in smart grids \cite{yang2013consensus}, and peer-to-peer census. \\

Typical distributed average consensus algorithms require agents to share their inputs with their neighbors \cite{aysal2009broadcast, benezit2010weighted, boyd2006randomized, jadbabaie2003coordination, olfati2007consensus, ren2005consensus, xiao2004fast}. This infringes upon agents' privacy, which is undesirable as certain agents maybe passively adversarial (a.k.a.~semi-honest~\cite{ben1988completeness} or honest-but-curious~\cite{manitara2013privacy}). Passive adversarial agents follow the prescribed consensus protocol honestly, but may try to use the information learned during an execution of the protocol to infer something about the inputs of other agents.\\

We present a distributed average consensus protocol that extends the privacy protocol proposed by Abbe et al., 2012~\cite{abbe2012privacy} for {\em complete} networks to the more general case of {\em incomplete} networks\footnote{A network is complete if and only if there is a dedicated communication channel between each pair of agents. An {\em incomplete} network is a network that is not complete.}. We show that the proposed protocol preserves the {\em privacy} of honest agents' inputs against colluding {passive} adversarial agents if the set of passive adversarial agents do not constitute a vertex cut in the network. Specifically, the passive adversarial agents collectively learn {\em nothing} about the inputs of the honest agents other than their average value. The latter is unavoidable, as it can be deduced from the global average whose computation is the purpose of an average consensus algorithm. \\

\subsection{\bfseries Prior Work}
While privacy can often be achieved by relying on generic completeness theorems for information-theoretic {\bf secure multi-party computation (MPC)}~\cite{ben1988completeness,chaum1988multiparty,Goldreich04}, those results however assume a complete network. There are few results in MPC for incomplete networks. Garay et al.~\cite{garay2008almost} studied secure computation in incomplete networks, and showed that arbitrary functions can be computed with information-theoretic privacy against $t$ colluding passive adversarial agents so long as the communication network is $(t+1)$-connected. However, their work relies on protocols for secure message transmission~\cite{dolev1993perfectly} to emulate pairwise private channels between every pair of agents. In addition to incurring a significant cost in terms of round- and message-complexity, relying on secure message transmission also requires the agents to have complete knowledge of the network topology. Our protocol adds minimal cost to existing distributed average consensus algorithms, and only requires agents to be aware of their neighbors. It is nevertheless interesting to note that our results also require $(t+1)$-connectivity to guarantee privacy against arbitrary subsets of $t$ colluding agents.\\

{\bf Differentially private protocols}, presented by Huang et al., 2012~\cite{huang2012differentially} and Nozari et al., 2017~\cite{nozari2017differentially}, can only compute an \emph{approximation} of the true average (rather than the exact average). Moreover, there is a trade-off between privacy and the achievable accuracy~\cite{nozari2017differentially}. On the other hand, our protocol does not perturb the average of the agents' inputs, and does not suffer from a privacy-accuracy trade-off.\\

Privacy protocols proposed by Manitara et al., 2013~\cite{manitara2013privacy}, Mo et al., 2017~\cite{mo2017privacy} overcome the trade-off in differentially private protocols by using random values or signals that vanish {\em eventually}. The privacy protocol by Wang, 2019~\cite{wang2019privacy} relies on random decomposition of inputs, instead of random input obfuscation. We note that these privacy protocols {\em cannot} preserve privacy of an honest agent's input {\em unless} the honest agent has an honest neighbor that has no passive adversarial neighbor~\cite{manitara2013privacy, mo2017privacy, wang2019privacy}. On the other hand, our protocol preserves the privacy of all the honest agents' inputs if every honest agent has an honest neighbor, i.e. the set of honest agents are not {\em cut} by the set of passive adversarial agents. We note that this condition on network connectivity is also sufficient for the privacy of agnets in protocols proposed by Altafini, 2019~\cite{altafini2019system}, and Rezazadeh, 2019~\cite{rezazadeh2019privacy}.\\

We note the {\bf \emph{observability}-based privacy protocols}, such as~\cite{alaeddini2017adaptive},~\cite{kia2015dynamic}, and~\cite{pequito2014design}, cannot protect privacy of an honest agent's input if that honest agent has a passive adversarial neighbor. On the other hand, our protocol can preserve privacy of the inputs of honest agents with passive adversarial neighbors, as long as the honest agents have an honest neighbor.\\

The privacy scheme by Gupta et al.~\cite{gupta2016confidentiality} assumes a centralized, trusted authority that distributes information to all agents each time they wish to run the consensus algorithm, and therefore is not distributed. \\




{\bf Homomorphic encryption-based protocols}, that use the Paillier cryptosystem~\cite{paillier1999public} to mask their inputs, can also achieve similar privacy guarantees as our protocol~\cite{hadjicostis2018privary, lazzeretti2014secure, ruan2019secure, yin2019accurate}. However, Paillier cryptosystem assumes finite computation power of a passive adversary. Whereas we consider an information-theoretic privacy, where the passive adversaries can have infinite computation power. \\

We also note that in some of the prior works, honest agents' inputs are deemed private if the passive adversarial agents are unable to compute the inputs {\em accurately}~\cite{alaeddini2017adaptive, altafini2019system, kia2015dynamic, manitara2013privacy, mo2017privacy, pequito2014design, rezazadeh2019privacy, wang2019privacy}. This is a weaker notion of privacy as there is no formal quantification for the privacy obtained. In this paper, we use the more standard notion of statistical distance for defining privacy in an information-theoretic manner. Formally, a distributed average consensus protocol is said to preserve the privacy of the honest agents' inputs if and only if the colluding passive adversarial agents are unable to learn anything about the inputs other than their sum (or average)~\cite{ben1988completeness, katz2014introduction}. 

\subsection{\bfseries Summary of Our Contributions}

We present a generalization of the privacy protocol proposed by Abbe et al., 2012~\cite{abbe2012privacy} for the case of incomplete networks. However, the protocol was independently proposed in our conference paper~\cite{gupta2017privacy}. The privacy guarantees presented in~\cite{gupta2017privacy} are however weaker than that in the current paper. We have proposed a modification of the protocol for the case when agents' inputs are finite real values in~\cite{gupta2019statistical}. In this paper, we consider the case when the agents' inputs are bounded integers.\\

The protocol constitutes two phases as summarized below. A detailed description is presented in Section~\ref{sec:pm}.

\begin{enumerate}
\setlength\itemsep{0.5em}
    \item 
In the first phase, each agent shares correlated random values with its neighbors, and computes a new ``effective input'' by adding the shared random values to the original input.

\item In the second phase, the agents run a non-private distributed average consensus protocol, such as {\em flooding} or any other protocol from the literature~\cite{boyd2006randomized, jadbabaie2003coordination, olfati2007consensus, xiao2004fast}, over their effective inputs computed in the first phase rather than the original inputs.
\end{enumerate}
~

The first phase is designed to ensure that the sum of the agents' effective inputs is equal to the sum of their original inputs under an appropriate modulo operation. Therefore, the above two-step process outputs the correct average value of the agents' original inputs. We show that the protocol preserves the privacy ---in a formal sense and under certain conditions, as discussed below---regardless of the average consensus protocol used in the second phase. 
We prove this by showing that privacy holds even if all the effective inputs of the honest agents are revealed to the colluding passive adversarial agents in the second phase.\\

Our notion of privacy is adopted from 
the literature of information-theoretic secure multi-party computation~\cite{Goldreich04}. Formally, the privacy requires that the entire \emph{view} (defined shortly) of a group of colluding agents throughout the execution of our protocol can be \emph{simulated} by those agents given (1)~their original inputs and (2)~the average of the original inputs of the honest agents (or, equivalently, the average of the original inputs of all the agents in the network). This should hold regardless of the true inputs of the honest agents.\\

In other words, the colluding passive adversarial agents learn nothing more about the collective inputs of the honest agents from an execution of the protocol other than the averages of the honest agents' inputs. Moreover, this holds regardless of any prior knowledge the adversarial agents may have about the inputs of (some of) the honest agents, or the distribution of those inputs. 
We prove that our protocol satisfies this notion of privacy if the set of colluding passive adversarial agents is not a vertex cut of the communication network. Alternately, our protocol satisfies the above notion of privacy for every subset of honest agents that is not {\em cut} by the set of passive adversarial agents. The privacy claims are presented formally in Section~\ref{sub:suff} and~\ref{sec:ext}.
\section{\bfseries Notation and Preliminaries}
\label{sec:not}

\def\Z{{\mathbb Z}}
\def\G{\mathcal{G}}
\def\V{\mathcal{V}}
\def\N{\mathcal{N}}
\def\E{\mathcal{E}}

We let 
$\mathbb{Z}$ denote the set of integers, and let $\Z_q$ denote the set of integers $\{0, \ldots, q-1\}$. 
For a finite set $S$, we let $\mnorm{S}$ denote its cardinality; for an integer $q$, we let $\mnorm{q}$ denote its absolute value. 
If $x$ is an $n$-dimensional vector, then $x_i$ denotes its $i$th element and $\sum_i x_i$ simply denotes the sum of all its elements (unless the range of $i$ is specifically mentioned). We use $1_n$ to denote the $n$-dimensional vector all of whose elements is~$1$.\\

\label{sub:gt}
A simple undirected graph is represented as $\mathcal{G} = \{\mathcal{V}, \,\mathcal{E} \}$ where
the nodes $\mathcal{V} \triangleq \{1,\ldots,n\}$ denote the agents, and there is an edge $\{i, j\} \in \mathcal{E}$ iff there is a direct communication channel between agents $i$ and~$j$.
We let $\N_i$ denote the set of neighbors of an agent $i \in \mathcal{V}$, i.e., $j \in \N_i$ if and only if $\{i,\,j\} \in \mathcal{E}$. Note that $i \not \in \N_i$ since $\mathcal{G}$ is a simple graph.
We say two agents $i, j$ are \emph{connected} if there is a path from~$i$ to~$j$; since we consider undirected graphs, this notion is symmetric. A graph $\mathcal{G}$ is \emph{connected} if every distinct pair of nodes is connected; note that a single-node graph is connected.\\

\begin{definition}{(Vertex cut)}
A set of nodes $\mathcal{V}_{cut} \subset \mathcal{V}$ is a \emph{vertex cut} of a graph $\mathcal{G}=\{\mathcal{V}, \mathcal{E}\}$ if removing the nodes in~$\mathcal{V}_{cut}$ (and the edges incident to those nodes) renders the resulting graph unconnected. In this case, we say that $\mathcal{V}_{cut}$ \emph{cuts}~$\mathcal{V}\setminus \mathcal{V}_{cut}$.
\end{definition} 
A graph is \emph{$k$-connected} if the smallest vertex cut of the graph contains~$k$ nodes.\\

Let $\G =\{\V, \E\}$ be a graph. The \emph{subgraph induced by $\V' \subset \V$} is the graph $\G' = \{\V', \E'\}$ where $\E' \subset \E$ is the set of edges entirely within $\V'$ (i.e., $\E' = \{\{i, j\} \in \E \mid i, j \in \V'\}$). 
A graph $\mathcal{G}=\{\mathcal{V}, \mathcal{E}\}$ has \emph{$c$ connected components} if its vertex set $\V$ can be partitioned into disjoint sets $\V_1, \ldots, \V_c$ such that; (1)~$\G$ has no edges between $\V_i$ and $\V_j$ for $i \neq j$, and (2)~for all~$i$, the subgraph induced by $\V_i$ is connected.
Clearly, if $\mathcal{G}$ is connected then it has one connected component.\\

For a graph $\G=\{\V, \E\}$, we define its \emph{incidence matrix}
$\nabla \in \{-1,0,1\}^{\mnorm{\mathcal{V}}\times \mnorm{\mathcal{E}}}$
to be the matrix with $|\V|$ rows and $|\E|$ columns,
\[
	\nabla_{i,\,e} = \left\{\begin{array}{cl}1 & \hspace*{3pt} \text{if } e = \{i,\,j\} \text{ and } i < j\\ -1 &  \hspace*{3pt} \text{if } e = \{i,\,j\} \text{ and } i > j \\ 0 &  \hspace*{3pt} \text{otherwise.}\end{array}\right. 
\]
Note that $1_n^T\cdot \nabla = 0$. We rely on the following result:\\

\noindent \fbox{\begin{minipage}{0.47\textwidth}
\begin{lemma}
\label{lem:o_m}\cite[Theorem 8.3.1]{godsil2001algebraic}
Let $\G$ be an $n$-node graph with incidence matrix $\nabla$. Then
$\text{rank}(\nabla) = n-c$, where $c$ is the number of connected components of~$\mathcal{G}$. 
\end{lemma}
\end{minipage}}

\subsection{\bfseries Problem Formulation}
\label{sec:prob_f}
We consider a network of $n$ agents where the communication network between agents is represented by an undirected, simple graph~$\G=\{\V, \E\}$; that  is, agents $i$ and $j$ have a direct communication link between them if and only if $\{i, j\} \in \E$. \\

Each agent $i$ holds a (private) input~$s_i \in \Z_q = \{0, \ldots, q-1\}$ for some publicly known, integer bound~$q>1$.\footnote{The proposed protocol can be easily extended for negative inputs. Suppose that the input of an agent $i$, let it be denoted by $x_i$, belongs to $\{q_1, \ldots, q_2\}$, where $q_1 \leq q_2 \in \Z$ are known. Then, we have $s_i = x_i - q_1 \in \{0, \ldots, q_2 - q_1\}$, and the average of $\{x_i\}$ is given by $\sum_i x_i /n = \sum_i s_i / n + q_i$.} Throughout the paper we will assume that the value of the total number of agents $n$ and the value of the upper bound $q$ on the agents' inputs is known to all the agents. \\

\def\C{\mathcal{C}}
\def\H{\mathcal{H}}
We let $s = [s_1,\ldots,\, s_n]^T$. 
A distributed average consensus algorithm is an interactive protocol allowing the agents in the network to each compute the average of the agents' inputs, i.e., after execution of the protocol each agent outputs the value $\bar s = \frac{1}{n} \, \sum_i s_i$. We consider a distributed average consensus protocol that ensures privacy of agents against some fraction of \emph{passive} adversarial agents in the network. \\

We let $\C \subset \V$ denote the set of adversarial agents, and let $\H = \V\setminus \C$ denote the remaining honest agents.\\

\begin{definition}
\label{def:v}
The \emph{View} of adversarial agents in $\C$ is the information constituting the inputs, internal states and received protocol messages of all the agents in $\C$ during an execution of the protocol. 
\end{definition}
~

Privacy requires that the entire \emph{view} of the adversarial agents does not provide any information about the inputs of honest agents other than the sum of their inputs, which is unavoidable if the privacy protocol does not affect the accuracy of the average value of the inputs (which is the case here) and all the agents (including adversarial agents) learn the value of $\sum_i s_i = n \bar s$ (assuming $n$ is known apriori to all the agents). This privacy definition is formalized below.\\



\def\view{{\sf View}}

Let $s_\C$ denote a set of inputs held by the adversarial agents, and $s_\H$ a set of inputs held by the honest agents. Fixing some protocol, we define $\view_\C(s)$ as follows:\\


\begin{definition}\label{def:view}
$\view_\C(s)$ is a random variable denoting the view of the adversarial agents $\C$ in an execution of the distributed average consensus protocol when all the agents begin holding inputs~$s$.
\end{definition}
~

Then, our privacy definition requires that the statistical distance between the prior and posterior probability distribution of the honest agents' inputs is zero in the {\empty view} of the adversary. Specifically, we have the following information-theoretic definition of privacy, which is borrowed from the literature on information-theoretic secure multiparty computation~\cite{ben1988completeness}.\\

\begin{definition}\label{def:ip}
A distributed average consensus protocol is \emph{(perfectly) $\C$-private} if for all $s, s' \in \Z^n_q$ such that  $s_\C=s'_\C$ and 
$\sum_{i \in \H} s_i = \sum_{i \in \H} s'_i$,
the distributions of $\view_\C(s)$ and $\view_\C(s')$ are identical.
\end{definition}
~

We remark that the privacy definition makes sense even if $|\C|=n-1$, though in that case the definition is vacuous since $s_\H = \sum_{i \in \H} s_i$ and so revealing the sum of the honest agents' inputs reveals the (single) honest agent's input.\\

The above privacy definition equivalently states that for any distribution $S$ (known to the colluding adversarial agents) over the honest agents' inputs, the distribution of the honest agents' inputs conditioned on the adversarial agents' \emph{view} is identical to the distribution of the honest agents' inputs conditioned on their sum.\\



In the subsequent section, we present the details of our privacy protocol and present the formal privacy guarantees. 

\section{\bfseries Privacy Protocol}
\label{sec:pm}

As described previously, our protocol has a two-phase structure. In the first phase, each agent~$i$ computes an ``effective input'' $\tilde s_i$ based on its original input~$s_i$ and random values it sends to its neighbors; this is done while ensuring that $\sum_i \tilde s_i \bmod p$ is equal to $\sum_i s_i$ for some publicly known integer~$p$ (see below). In the second phase, the agents use any (non-private) distributed average consensus protocol $\Pi$ to compute the average of $\{n\tilde s_i\}$ or equivalently $\sum_i \tilde s_i$, reduce that result modulo~$p$, and then divide by~$n$. This gives the correct average $\frac{1}{n} \, \sum_i s_i$, and thus all that remains is to analyze the privacy. \\

It may at first seem strange that we can prove privacy of our protocol without knowing anything about the distributed average consensus algorithm $\Pi$ used in the second phase of our algorithm. We do this by making a ``worst-case'' assumption about $\Pi$, namely, that it simply reveals all the agents' inputs to all the agents. Such an algorithm is, of course, not at all private; for our purposes, however, this does not immediately violate privacy because $\Pi$ is run on the agents' scaled \emph{effective} inputs~$\{n\tilde{s}_i\}$ rather than their true inputs~$\{s_i\}$. We make the following additional assumption:\\
\begin{itemize}
\item[\textbf{A1:}] The communication links between the agents are private during the first phase of our protocol\footnote{Alternately, private communication can be ensured using standard cryptographic techniques~\cite{katz2014introduction}.}. That is, if agent $i$ sends a value over link $\{i, \, j\}$ then only agent $j$ can retrieve that value.\\
\end{itemize}

Note that the communication links need not be private in the second phase of the protocol. This saves significant computation and communication cost as cryptographic techniques for ensuring privacy over communication links have high computation and communication costs.\\

Under Assumption~\textbf{A1}, the view of the adversarial agents consist of the initial inputs of the agents in $\C$, their internal states and all the protocol messages they receive during an execution of the first phase of our protocol, and the vector~$\tilde{s} = [\tilde{s}_1, \ldots, \tilde{s}_n]^T$ of all agents' effective inputs at the end of the first phase. The definition of privacy (cf.\ Definition~\ref{def:ip}) remains unchanged.\\

Before continuing with the analysis of privacy, we describe our first-phase algorithm. \\

\noindent \fbox{\begin{minipage}{0.47\textwidth}
Let $p$ be an integer such that $p > n\cdot (q-1) \geq \sum_i s_i$. The \textbf{first phase} of our protocol proceeds as follows:
\begin{enumerate}
	\item Each agent $i \in \mathcal{V}$ chooses independent, uniform values $r_{ij} \in \Z_p$ for all $j \in \mathcal{N}_i$, and sends $r_{ij}$ to agent~$j$.
	\item Each agent $i \in \mathcal{V}$ computes a mask $a_i \in \Z_p$ as,
		\begin{align}
			a_i = \sum_{j \in \N_i}(r_{ji}-r_{ij}) \bmod p, \label{eqn:i_vn}
		\end{align}
	\item Each agent $i \in \mathcal{V}$ computes  effective input
	\begin{align}
		\tilde{s}_i = (s_i + a_i) \bmod p. \label{eqn:mask_coeff}
	\end{align}
\end{enumerate}
\end{minipage}}
~\\

\noindent{\bf Correctness:} From~\eqref{eqn:mask_coeff},
\[
\sum_i \tilde s_i  =  \sum_i s_i + \sum_i a_i \bmod p.
\]
When $\G$ is undirected then
\begin{align*}
\sum_i a_i  =  \sum_i \sum_{j \in N_i}(r_{ji}-r_{ij}) = 0 \mod p,
\end{align*}
Thus, 
$\sum_i \tilde s_i = \sum_i s_i \bmod p$. Since $\sum_i s_i < p$ by choice of~$p$, this implies that $\sum_i \tilde s_i \bmod p$ is equal to $\sum_i s_i$ over the integers. Hence, when $\G$ is undirected then the correctness of our overall algorithm (i.e., including the second phase) follows from above.

\begin{figure}[htb!]
\begin{center}
	\includegraphics[width=0.45\textwidth]{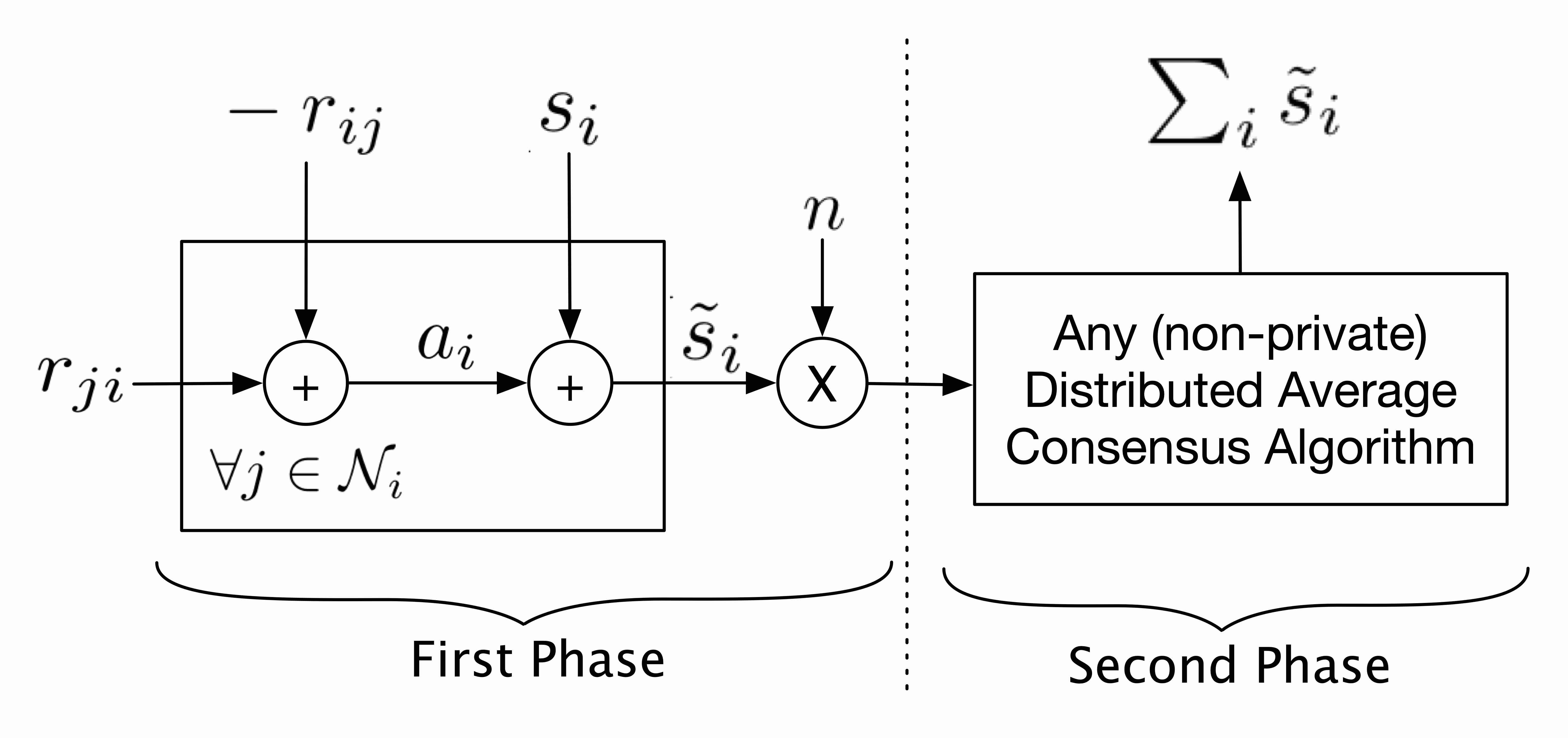}    
	\caption{\footnotesize{Schematic of the protocol as viewed by an agent $i$.}}
	\label{fig:prot}
\end{center}
\end{figure}

\subsection{\bfseries Privacy Guarantee}
\label{sec:pa}

We show here that $\C$-privacy holds as long as $\C$ is not a vertex cut of~$\G$. For an edge $e=\{i,j\}$ in the graph with $i<j$,
define
\begin{align*}
    b_e = r_{ji}-r_{ij} \bmod p
\end{align*}
Note that for any two independent random variables $x$ and $y$ in $\mathbb{Z}_p$, if at least one of them is uniformly distributed in $\mathbb{Z}_p$ then the random variable $ = x + y \bmod p$ is also uniformly distributed in $\mathbb{Z}_p$~\cite{katz2014introduction}. Therefore, since the random variables $r_{ij}$ and $r_{ji}$ are independently and uniformly distributed in~$\Z_p$, the random variable $b_e$ is also uniformly distributed in $\Z_p$, for every edge $e = \{i, \, j\}$.\\

Let $b=[b_{e_1}, \ldots]$ be the collection of $b_e$'s for all the edges in~$\G$. Let $a=[a_1, \ldots, a_n]^T$ denote the masks used by the agents in the first phase. Then, 
\[a = \nabla \, b \mod p.\]
Thus, $a$ is uniformly distributed over the vectors in the span of the columns of~$\nabla$ in $\Z_p$, which we denote by the set
\[L(\nabla) = \{\nabla \, b \, \bmod p \, | \, b \in \Z^{\mnorm{\E}}_p\}.\]
The following is proved using the fact that 
$\text{rank}(\nabla)=n-1$ when $\G$ is connected (cf.\ Lemma~\ref{lem:o_m}):\\

\begin{lemma}
\label{lem:dist_a}
If $\mathcal{G}$ is an undirected connected graph then $a$ is uniformly distributed over $\Z_p^n$ subject to the constraint that $\sum_i a_i = 0 \bmod p$. 
\end{lemma}
~

Formal proof of Lemma~\ref{lem:dist_a} is presented in Appendix~\ref{sub:dist}.
Since $\tilde s_i = s_i + a_i \bmod p$, we have\\

\begin{lemma}
\label{lem:cond_mask}
	If $\mathcal{G}$ is an undirected connected graph then given the value of $s \in \Z^n_q$ the effective inputs $\tilde s$ are uniformly distributed in $\Z_p^n$ subject to the constraint: $\sum_i \tilde s_i = \sum_i s_i \bmod p$.
\end{lemma}
~

Formal proof of Lemma~\ref{lem:cond_mask} is presented in Appendix~\ref{sub:cond}. Lemma~\ref{lem:cond_mask} implies privacy for the case when $\C=\emptyset$, i.e., when there are no adversarial agents. In that case, the view of the adversary consists only of the effective inputs~$\tilde{s}$, and Lemma~\ref{lem:cond_mask} shows that the distribution of those values
depends only on the sum of the agents' true inputs. 
Below, we extend this line of argument to the case of nonempty~$\C$.\\




\label{sub:suff}

Fix some set $\C$ of adversarial agents, and recall that $\H=\V\setminus \C$. Let $\E_\C$ denote the set of edges incident to~$\C$, and let $\E_{\H} = \E \setminus \E_\C$ be the edges incident only to honest agents. 
Note that now the adversarial agents' \emph{view} contains (information that allows it to compute) $\{b_e\}_{e \in \E_\C}$ in addition to the honest agents' effective inputs~$\{\tilde s_i\}_{i \in \H}$.\\

The key observation enabling a proof of privacy is that the values $\{b_e\}_{e \in \E_{\H}}$ are uniform and independent in $\Z_p$ \emph{even conditioned on the values of~$\{b_e\}_{e \in \E_\C}$}. Thus, as long as $\C$ is not a vertex cut of~$\G$, an argument as earlier implies that the masks $\{a_i\}_{i \in \H}$ are uniformly distributed in $\Z_p^{|\H|}$ subject to 
\[\sum_{i \in \H} a_i = -\sum_{i \in \C} a_i  = - \sum_{i \in \C} \left( \sum_{e \in \E_\C} \nabla_{i,e} \, b_e  \right) \mod p\]
given the values~$\{b_e\}_{e \in \E_\C}$. Therefore, for the adversarial agents in $\C$ the effective inputs $\{\tilde s_i\}_{i \in \H}$ are uniformly distributed in $\Z_p^{|\H|}$ subject to 
\begin{align}
    \sum_{i \in \H} \tilde s_i = \sum_{i \in \H} s_i - \sum_{i \in \C} a_i = \sum_{i \in \V} s_i - \sum_{i \in \C} \tilde s_i \mod p \label{eqn:tilde_s}
\end{align}
From the above observations, we obtain the following theorem. The formal proof of the theorem below is deferred to Appendix~\ref{sub:perf}.\\

\begin{theorem}
\label{thm:perf}
Suppose that $\G$ is an undirected connected graph and Assumption~\textbf{A1} holds true. If $\C$ is not a vertex cut of $\G$ then the distributed average consensus protocol presented in Section~\ref{sec:pm} is $\C$-private.
\end{theorem}
~

According to Theorem~\ref{thm:perf}, the protocol in Section~\ref{sec:pm} preserves privacy of all the honest agents' inputs if every honest agent has an honest neighbor in the underlying communication network. As a corollary, we have\\

\begin{corollary}
\label{cor:f}
Suppose that $\G$ is an undirected connected graph and Assumption~\textbf{A1} holds true. If $\G$ is $(t+1)$-connected then for every $\C$ with $|\C| \leq t$ the distributed average consensus protocol presented in Section~\ref{sec:pm} is $\C$-private.
\end{corollary}



\section{\bfseries Illustration}
\label{sec:illus}
\begin{figure}[htp!]
\begin{center}
	\includegraphics[width=0.25\textwidth]{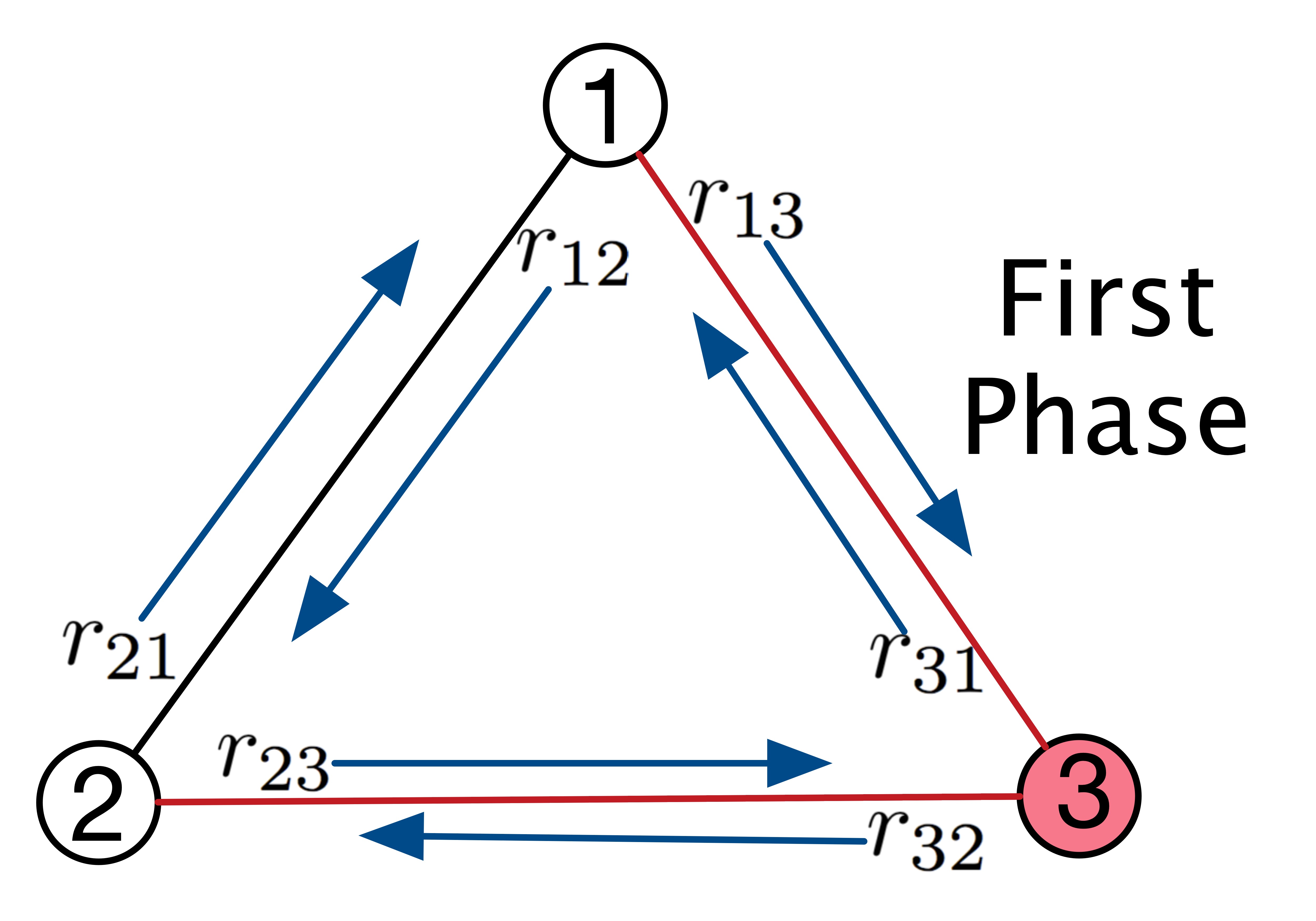}    
	\caption{\footnotesize{Illustration of the first phase of the protocol. The arrows represent the flow of information.}}
	\label{fig:illust}
\end{center}
\end{figure}

In this section, we demonstrate the distributed average consensus protocol on a simple network of $3$ agents; $\mathcal{V} = \{1,\, 2, \, 3\}$ and $\mathcal{E} = \left\{ \{1,\,2\}, \, \{1,\,3\}, \, \{2,\,3\} \right\}$, as shown in Fig. \ref{fig:illust}. Let the values of $q$ and $p$ be $10$ and $30$, respectively. Let, $s_1 = 4$, $s_2 = 7$ and $s_3 = 3$. In the first phase:
\begin{enumerate}
        \item 
        As shown in Fig. \ref{fig:illust}, all pair of adjacent agents $i$ and $j$ exchange the respective values of $r_{ij}$ and $r_{ji}$ (chosen independently and uniformly in $\mathbb{Z}_{p}$) with each other. Consider a particular instance where  
        \begin{align*}
            [r_{12}, r_{21}, r_{23}, r_{32},r_{31}, r_{13}] = [14,11,17,5,3,8]
        \end{align*}
        \item
            The agents compute their respective masks, 
            \[a_1 =  \left((r_{21} - r_{12}) + (r_{31} - r_{13}) \right) \bmod p = 22\]
            and similarly, $a_2 = 21$ and $a_3 = 17$. It is easy to verify that $(a_1 + a_2 + a_3) \bmod 30 = 0$.
        \item 
            The agents compute their respective effective inputs,
            \[\tilde{s}_1 = (s_1 + a_1) \bmod p  = (4 + 22) \bmod 30 = 26\]
            and similarly, $\tilde{s}_2 = 28$ and $\tilde{s}_3 = 20$.
\end{enumerate}

After the first phase, each agent uses a (non-private) distributed average consensus protocol $\Pi$ in the second phase to compute $\sum_{i}\tilde{s}_i$. It is easy to verify that 
\[\sum_{i}\tilde{s}_i \bmod 30 = 14 = \sum_{i} s_i = 14\]
Let $\C = \{3\}$ and so, $\E_{\C} = \{\{1,\,3\}, \, \{2,\,3\}\}$. As Agent $3$ does not cut the graph, according to Theorem~\ref{thm:perf}, the only information Agent $3$ can learn about inputs $s_1$ and $s_2$ is that \\$s_1 + s_2 = \tilde{s}_1 + \tilde{s}_2 + a_3 \bmod p = (26 + 28 + 17) \bmod 30 = 11$.

\section{\bfseries Extended Privacy}
\label{sec:ext}

In this section, we present an extension of Theorem~\ref{thm:perf} for the case when $\C$ is a vertex cut, by relaxing our definition of $\C$-privacy to $(\C, \, \H)$-privacy as follows, where $\H$ is a subset of $\V \setminus \C$. Apart from $\H$, the notation remains unchanged.\\

\begin{definition}\label{def:ip_2}
A distributed average consensus protocol is \emph{(perfectly) $(\C, \, \H)$-private} if for all $s, s' \in \Z^{\mnorm{\H}}_q$ subject to $s_{\V \setminus \H}=s'_{\V \setminus \H}$ and 
$\sum_{i \in \H} s_i = \sum_{i \in \H} s'_i$,
the distributions of $\view_\C(s)$ and $\view_\C(s')$ are identical.
\end{definition}
~


If a protocol is $(\C, \, \H)$-private, for a subset $\H \subset \V \setminus \C$, then the set of adversarial agents $\C$ can learn nothing about the inputs of agents in $\H$ other than their sum. Note that $\C$-privacy is equivalent $(\C, \, \V \setminus \C)$-privacy. Therefore, if a distributed average consensus protocol is $\C$-private then it is $(\C,\, \H)$-private for all $\H \subseteq \V \setminus \C$. This is the reason why Definition \ref{def:ip_2} is a relaxation of Definition \ref{def:ip}. As a corollary of Theorem~\ref{thm:perf}, we have\\

\begin{corollary}
\label{cor:ext_res}
If Assumptions \textbf{(A1)}-\textbf{(A4)} hold and $\C$ does not \emph{cut} $\H$, then the distributed average consensus protocol presented in Section~\ref{sec:pm} is $(\C, \, \H)$-private.
\end{corollary}
~

The above privacy claim is illustrated below in Fig.~\ref{fig:gen_conn}.

\begin{figure}[htb!]
\begin{center}
	\includegraphics[width=0.35\textwidth]{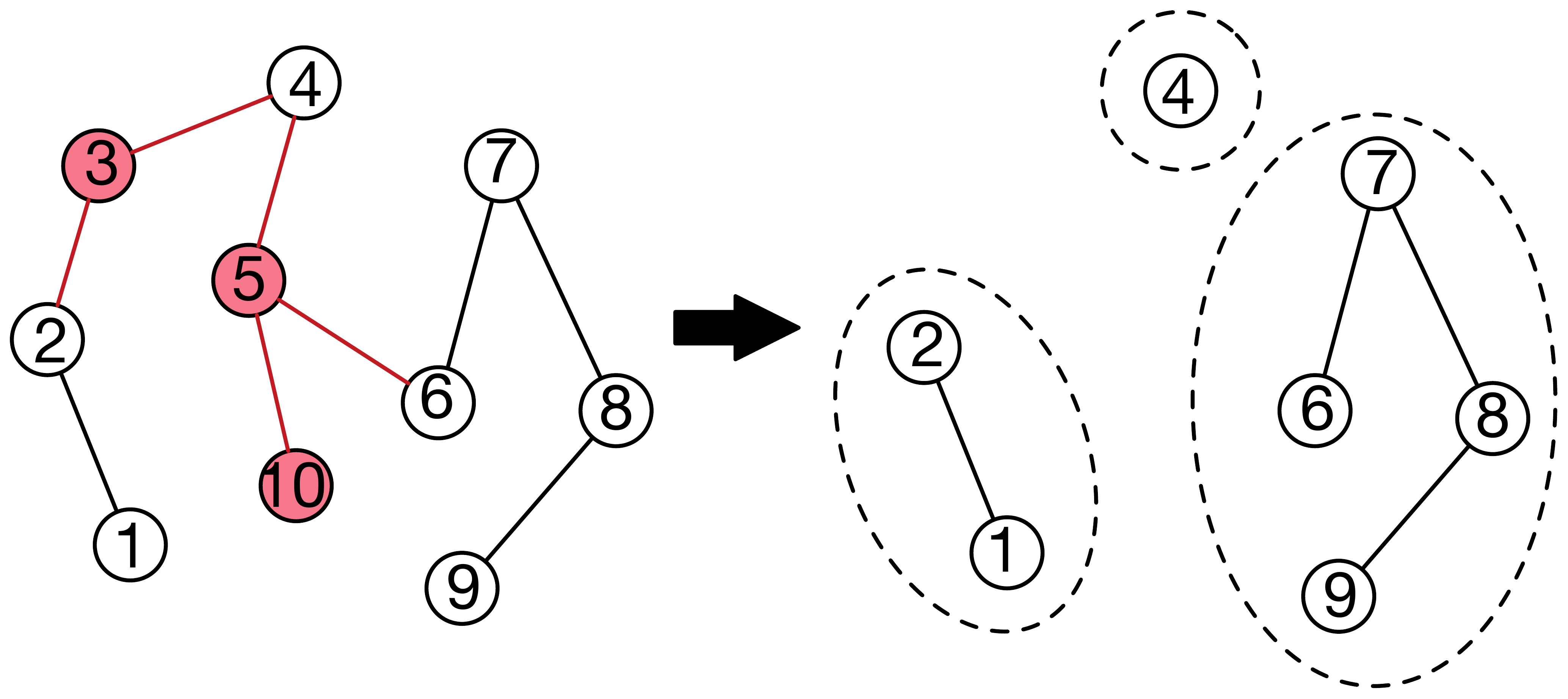}    
	\caption{\footnotesize{In this case, the adversarial agents $\C = \{3, \, 5, \, 10\}$ cut the graph into $3$ connected components with set of agents $\H_{1} = \{1,2\}$, $\H_{2} = \{4\}$ and $\H_3 = \{6,7,8,9\}$ (and edges incident to the respective honest agents). The distributed average consensus protocol presented in Section~\ref{sec:pm} preserves the privacy of honest agents' inputs in each group $\H_{i}, \, i = 1,\,2, \,3$, in the formal sense as defined in Definition \ref{def:ip_2}. However, as $\sum_{i \in \H_2} s_i = s_4$, the value of $s_4$ is revealed to the adversarial agents. }}
	\label{fig:gen_conn}
\end{center}
\end{figure}

\section{\bfseries Summary}
\label{sec:dis}

This paper presents a private distributed average consensus protocol that preserves information-theoretic privacy of honest agents' inputs against a set of passive adversarial agents that do not form a vertex cut in the underlying communication network. In other words, the presented protocol preserves privacy of all the honest agents' inputs if every honest agent has an honest neighbor. This implies that privacy of honest agents' inputs is preserved against at most $t$ arbitrary colluding passive adversarial agents if the network has $(t+1)$-connectivity. The privacy requires that the colluding passive adversarial agents learn nothing about the inputs of honest agents beyond their sum (or average). The latter is unavoidable, as it can be deduced from the global average whose computation is the purpose of running the consensus algorithm. Upon generalizing the above privacy for subsets of honest agents, the protocol is shown to preserve the privacy of the collective inputs of a subset of honest agents as long as every agent in the subset has an honest neighbor.
\bibliographystyle{plain}
{\scriptsize\bibliography{references_consensus}}

\appendix
\subsection{\bfseries Proof of Lemma \ref{lem:dist_a}}
\label{sub:dist}
The proof is obvious for $n = 1$. Henceforth, $n > 1$. Let $\nabla_{*,e}$ to denote the column of $\nabla$ corresponding to the edge $e \in \mathcal{E}$. Let $\E'$ be a subset of $\E$ with $n-1$ edges such that the columns $\{\nabla_{*,e} \, | \, e \in \E'\}$ of the oriented incidence matrix are linearly independent. Note that, due to Lemma~\ref{lem:o_m}, there exists such a subset $\E'$ when $\G$ is connected. 
Let,
\[a' =  \sum_{e \in \E'}\nabla_{*,e} \, b_e  \mod p\]
Let, $\G' = \{\G, \, \E'\}$ and $\nabla'$ be the incidence matrix of $\G'$ whose columns are given by $\{\nabla_{*,e} \, | \, e \in \E'\}$.  Then,
\[a' = \nabla' \, b \mod p\]
As non-zero elements of $\{\nabla'_{*,e}\}_{e \in \E}$ belong to $\{-1, \, 1\}$ and $\{\nabla_{*,e} \, | \, e \in \E'\}$ are linearly independent, thus $\sum_{e \in \E'}\nabla'_{*,e} \, b_e  \mod p = 0$ if and only if $b_e = 0, \, \forall e \in \E'$. Therefore, for two set of values $\{b^1_e, \, | \, e \in \E'\}$ and $\{b^2_e, \, | \, e \in \E'\}$,
\[\sum_{e \in \E'}\nabla'_{*,e} \, b^1_e = \sum_{e \in \E'}\nabla'_{*,e} \, b^2_e \mod p \] 
if and only if $b^1 = b^2_e, \, \forall e \in \E'$. Thus, every value of $\{b_e\}_{e \in \E'}$ generates a unique value of $a'$. \\

As $\{b_e\}_{e \in \E'}$ are uniformly distributed in $\mathbb{Z}^{n-1}_p$, the above implies that $a'$ is uniformly distributed over all $p^{n-1}$ points in 
\[L(\nabla') = \{\nabla' \, b \, \bmod p \, | \, b \in \mathbb{Z}_p^{n-1}\},\] 
We show that $a$ is also uniformly distributed in $L(\nabla')$ using reasoning from induction as follows. \\

\noindent For any integer $0 \leq k < |\E|-|\E'|$, let
\begin{align*}
    a^{(k)} = \sum_{e \in \E^{(k)}}\nabla_{*,e} \, b_e  \mod p
\end{align*}
where $\E^{(k)}$ is the set of edges generated by adding any $k$ edges from $\E \setminus \E'$ in $\E'$. Clearly, $a^{0} = a'$, which as shown above is uniformly distributed over all $p^{n-1}$ points in $L(\nabla')$. Now, we show that if $a^{(k)}$ is uniformly distributed over all $p^{n-1}$ points in $L(\nabla')$ for some $k$ then the same is true for $a^{(k+1)}$.\\
Let $e^{(k+1)}$ be an edge in $\E \setminus \E^{(k+1)}$ and 
\begin{align*}
    a^{(k+1)} = a^{(k)} + \nabla_{*,e^{(k)}}\, b_{e^{(k+1)}} \mod p.
\end{align*}
As $a^{(k)}$ is assumed to be uniformly distributed over $L(\nabla')$, we can substitute $a^{(k)}$ in equation above by 
\[\sum_{e \in \E'}\nabla_{*,e}\, b_e \, \bmod p\]
where $b_e \in \Z \, \forall \, e \in \E'$. This implies, 
\begin{align}
    a^{(k+1)} = \sum_{e \in \E'}\nabla_{*,e} \, b_e + \nabla_{*,e^{(k)}}\, b_{e^{(k+1)}} \mod p \label{eqn:a_p}
\end{align}
As $\G'$ is connected, there exists a path in $\G'$ between the terminal nodes of the edge $e^{(k+1)}$. Therefore, there exists $\mu_e \in \{-1, 0, 1\}$ for all $e \in \E'$, such that 
\begin{align}
    \nabla_{*,e^{(k+1)}} = \sum_{e \in \E'} \nabla_{*,e} \, \mu_e \label{ieqn:nabla_p}
\end{align}
Substituting~\eqref{ieqn:nabla_p} in~\eqref{eqn:a_p}, we obtain,
\begin{align}
    a^{(k+1)} = \sum_{e \in \E'}\nabla_{*,e} (b_e + \mu_e b_{e^{(k+1)}}) \mod p \label{eqn:final_a_p}
\end{align}
As~$\{b_e\}_{e \in \E'}$ is uniformly distributed over all points in $\mathbb{Z}^{n-1}_p$ and $b_{e^{(k+1)}}$ is independent from all $\{b_e\}_{e \in \E'}$, $\{b_e + \mu_e b_{e^{(k+1)}} \bmod p\}_{e \in \E'}$ is uniformly distributed over all the points in $\mathbb{Z}^{n-1}_p$. Therefore,~\eqref{eqn:final_a_p} implies that $a^{(k+1)}$ is uniformly distributed over all $p^{n-1}$ points in $L(\nabla')$.\\

As $a^{(k+1)} = a$ when $k = |\E|-|\E'|-1$, reasoning from induction implies that $a$ is uniformly distributed over all $p^{n-1}$ points in $L(\nabla')$.\\

As $1_n^T \, a = 0 \bmod p$ as $1_n^T \nabla = 0_{|\E|}^T$ when $\G$ is connected, the above implies that $a$ is uniformly distributed over $\Z_p^n$ subject to the constraint: $\sum_i a_i = 0 \bmod p$.

\subsection{\bfseries Proof of Lemma \ref{lem:cond_mask}}
\label{sub:cond}
Since $\tilde{s}_i = s_i + a_i \bmod p$, and $s_i, a_i$ are independent random variables, we get
\begin{align*}
    &Pr\left(\tilde{s} | s \right) = Pr\left(a = (\tilde{s} - s) \bmod p\right)
\end{align*}
From Lemma~\ref{lem:dist_a} we know that
\begin{align*}
	Pr(a) = \left\{ \begin{array}{ccc} 1/p^{n-1} &, & \sum_i a_i = 0 \bmod p \\ 0 &, & \text{otherwise} \end{array}\right.
\end{align*}
when $\G$ is connected. Therefore, 
\begin{align*}
	Pr(\tilde s | s) = \left\{ \begin{array}{ccc} 1/p^{n-1} &, & \sum_i \tilde{s}_i = \sum_i s_i \bmod p \\ 0 &, & \text{otherwise} \end{array}\right.
\end{align*}
when $\G$ is connected. For a given value of $s$, there can be at most $p^{n-1}$ values of $\tilde{s}$ that satisfy $\sum_i \tilde{s}_i = \sum_i s_i \bmod p$. Thus, the above implies that $\tilde{s}$ are uniformly distributed in $\Z^n_p$ subject to $\sum_i \tilde{s}_i = \sum_i s_i \bmod p$ when $\G$ is connected.

\subsection{\bfseries Proof of Theorem \ref{thm:perf}}
\label{sub:perf}

Let $\G_{\H} = \{\H, \, \E_{\H}\}$ be the graph of honest agents (and edges incident to only honest agents) and $\nabla_{\H}$ be its \emph{incidence matrix}. Note that $\G_\H$ is undirected as $\G$ is undirected.\\
Due to Assumption~\textbf{A1}, 
\[\view_{\C}(s) = \left\{s_{\C}, \, \{ \tilde{s}_i \}, \, \{b_e\}_{e \in \E_{\C}}\right\}.\] 
Each $a_i$ can be decomposed as follows:
\begin{align*}
    a_i = \sum_{e \in \E_{\H}}\nabla_{i,e} \, b_e + \sum_{e \in \E_{\C}}\nabla_{i,e} \, b_e\mod p
\end{align*}
As the random values $\{b_e\}_{e \in \E_{\H}}$ are uniformly and independently distributed in $\mathbb{Z}_p$ (given the values $\{b_e\}_{e \in \E_{\C}})$, this implies that the collection of random vectors $\{ \sum_{e \in \E_{\H}}\nabla_{i,e} b_e \bmod p\}_{i \in \H}$ is uniformly distributed over $\mathbb{Z}^{|\H|}_p$ subject to the constraint 
\[\sum_{i \in \H}\left(\sum_{e \in \E_{\H}}\nabla_{i,e} b_e\right) = 0 \mod p\]
when $\G_{\H}$ is connected (cf. Lemma~\ref{lem:dist_a}). Thus, if $\G_{\H}$ is connected then
\begin{align*}
	&Pr(a_{\H} |  \{b_{e}\}_{ e \in \E_{\C}}) \\
	&= \left\{ \begin{array}{ccc} 1/p^{|\H|-1} &, & \sum_{i \in \H} a_i = -\sum_{i \in \C} a_i \bmod p \\ 0 & , & \text{otherwise}\end{array} \right.
\end{align*}
where, $a_i = \sum_{e \in \E_{\C}} \nabla_{i,e}  b_e \bmod p, \, \forall i \in \C$ and $a_{\H}$ denotes the vector of honest agents masks $\{a_i\}_{i \in \H}$.
Combining the above with the fact that $\tilde s_i = s_i + a_i \bmod p, \, \forall i$, where $s_i$ and $a_i$ are independent for all $i$, implies that ($\tilde{s}_{\H}$ is the vector of $\{\tilde s_i\}_{i \in \H}$) 
\begin{align}
    Pr\left(\tilde{s}_{\H} | s_{\H}, \, \{b_{e}\}_{ e \in \E_{\C}} \right) = 1/p^{|\H|-1} \label{eqn:f_cond}
\end{align}
for all the values $\tilde{s}_{\H}$ in $\mathbb{Z}^{|\H|}_p$ that satisfy 
\[\sum_{i \in \H} \tilde s_i = \sum_{i \in \H}s_i - \sum_{i \in \C}\sum_{e \in \E_{\C}}\nabla_{i,e}b_e \mod p\]
when $\G_{\H}$ is connected. 
As $\{b_e\}_{e \in \E}$ are independent to the inputs $\{s_i\}$, thus 
\[Pr\left(\tilde{s}_{\H} | s_{\H}, \, \{b_{e}\}_{ e \in \E_{\C}} \right) = \frac{Pr\left(\tilde{s}_{\H}, \, \{b_{e}\}_{ e \in \E_{\C}} | s_{\H} \right)}{Pr(\{b_{e}\}_{ e \in \E_{\C}})}\]
As $a_i = \sum_{e \in \E_{\C}}\nabla_{i,e}b_e \bmod p, \, \forall i \in \C$ and $\tilde{s}_i = s_i + a_i \bmod p, \, \forall i$, thus from \eqref{eqn:f_cond} we get
\begin{align*}
    Pr\left(\view_{\C}(s) \right) \equiv Pr\left(\view_{\C}(s')\right)
\end{align*}
for all inputs $s, s'$ in $\Z^{n}_q$ that satisfy $s_{\C} = s'_{\C}$ and $\sum_{i \in \V}s_i = \sum_{i \in \V}s'_i \bmod p$ when $\G_{\H}$ is connected.





\end{document}